\documentclass[10pt,journal,compsoc]{IEEEtran}
\pdfoutput=1

\ifCLASSOPTIONcompsoc
  \usepackage[nocompress]{cite}
\else
  \usepackage{cite}
\fi
\ifCLASSINFOpdf
\else
\fi

\usepackage{cite}
\usepackage{graphicx}
\usepackage{url} 

\usepackage{algorithm}
\usepackage{algorithmic}
\usepackage{amsmath}
\usepackage{amsfonts}
\usepackage{multirow}
\usepackage{color}
\usepackage{diagbox}
\usepackage{threeparttable}

\begin{document}
%
\title{Investigating EEG-Based Functional Connectivity Patterns for Multimodal Emotion Recognition}

\author{Xun~Wu,~
	Wei-Long~Zheng,~\IEEEmembership{Member,~IEEE,}
	and~Bao-Liang~Lu$^{*}$,~\IEEEmembership{Senior~Member,~IEEE}
	\IEEEcompsocitemizethanks{
		\IEEEcompsocthanksitem This work was supported in part by the National Key Research and Development Program of China (Grant 2017YFB1002501),  the National Natural Science Foundation of China (Grant No. 61673266  and No. 61976135), the Fundamental Research Funds for the Central Universities, and the 111 Project. 
		\IEEEcompsocthanksitem Xun Wu and Bao-Liang Lu are with the Center for Brain-Like Computing and Machine Intelligence, Department of Computer Science and Engineering, the Key Laboratory of	Shanghai Education Commission for Intelligent Interaction and Cognitive Engineering, the Brain Science and Technology Research Center, and Qing Yuan Research Institute, Shanghai Jiao Tong University, 800 Dong Chuan Road, Shanghai 200240, China.
		\IEEEcompsocthanksitem Wei-Long Zheng is with the Clinical Data Animation Center, Department of Neurology, Massachusetts General Hospital, Harvard Medical School, 55 Fruit Street, Boston, Massachusetts, USA.
	}
\thanks{$^{*}$Corresponding author: {\tt\small bllu@sjtu.edu.cn}}
}

\IEEEtitleabstractindextext{%
\begin{abstract}
Compared with the rich studies on the motor brain-computer interface (BCI), the recently emerging affective BCI presents distinct challenges since the brain functional connectivity networks involving emotion are not well investigated. Previous studies on emotion recognition based on electroencephalography (EEG) signals mainly rely on single-channel-based feature extraction methods. In this paper, we propose a novel emotion-relevant critical subnetwork selection algorithm and investigate three EEG functional connectivity network features: strength, clustering coefficient, and eigenvector centrality. The discrimination ability of the EEG connectivity features in emotion recognition is evaluated on three public emotion EEG datasets: SEED, SEED-V, and DEAP. The strength feature achieves the best classification performance and outperforms the state-of-the-art differential entropy feature based on single-channel analysis for the EEG signals. The experimental results reveal that distinct functional connectivity patterns are exhibited for the five emotions of disgust, fear, sadness, happiness, and neutrality. Furthermore, we construct a multimodal emotion recognition model by combining the functional connectivity features from EEG and the features from eye movements or physiological signals using deep canonical correlation analysis. The classification accuracies of multimodal emotion recognition are $95.08\pm6.42\%$ on the SEED dataset, $84.51\pm5.11\%$ on the SEED-V dataset, and $85.34\pm2.90\%$ and $86.61\pm3.76\%$ for arousal and valence on the DEAP dataset, respectively. The results demonstrate the complementary representation properties of the EEG functional connectivity network features with eye movement data. In addition, we find that the brain networks constructed with fewer channels, i.e., 18 channels, achieve comparable performance with that of the 62-channel network with respect to multimodal emotion recognition and enable easier setups for BCI systems in real scenarios.
\end{abstract}

}

\maketitle

\IEEEdisplaynontitleabstractindextext

%
\IEEEpeerreviewmaketitle

\IEEEraisesectionheading{\section{Introduction}\label{sec:introduction}}

\IEEEPARstart{E}{motion} plays a crucial role in many aspects of our daily lives, such as social communication and decision-making. According to the Gartner hype cycle in 2019 \cite{gartner2019}, emotion artificial intelligence (AI) is one of the 21 emerging technologies that will significantly impact our society over the next 5 to 10 years. Emotion AI, also known as artificial emotional intelligence or affective computing \cite{poria2017review}, aims at enabling machines to offer the capabilities to recognize, understand, and process emotions. Compared with the rich studies on the motor brain-computer interface (BCI), the recently emerging affective BCI (aBCI) \cite{muhl2014survey} faces distinct challenges since the brain functional connectivity networks involving emotions are not well investigated \cite{Shanechi2019nn}. The aBCI technology aims to advance the human-computer interaction systems with the assistance of various devices to detect the affective states from neurophysiological signals. Therefore, the major challenge facing emotion AI and aBCI in the primary stage lies in emotion recognition \cite{thanapattheerakul2018emotion}.


In recent years, extensive endeavors have been devoted to emotion recognition. Among a variety of emotion recognition approaches, the modalities used to detect affective states primarily comprise two categories: the external behavioral signals, including facial expression \cite{ko2018brief}, speech \cite{schuller2018speech}, body language, etc., and the internal physiological signals \cite{shu2018review}, containing electroencephalography (EEG) \cite{alarcao2017emotions}, electrocardiography (ECG) \cite{xiefeng2019heart}, respiration, galvanic skin response, etc. These two categories have their own prominent properties. The external behavioral signals outperform in terms of convenience of data collection, while the physiological signals are believed to be more objective and reliable in conveying emotions. As a result, multimodal emotion recognition has become the major trend, since it may leverage the complementary representation properties of different modalities. Nevertheless, most existing studies have focused on the fusion of visual and audio signals \cite{poria2016convolutional} \cite{zhang2016enhanced}, while few studies have combined the behaviors with physiological signals \cite{povolny2016multimodal} \cite{soleymani2011multimodal}. 

\begin{figure*}[t!]
	\centering\includegraphics[scale=0.9]{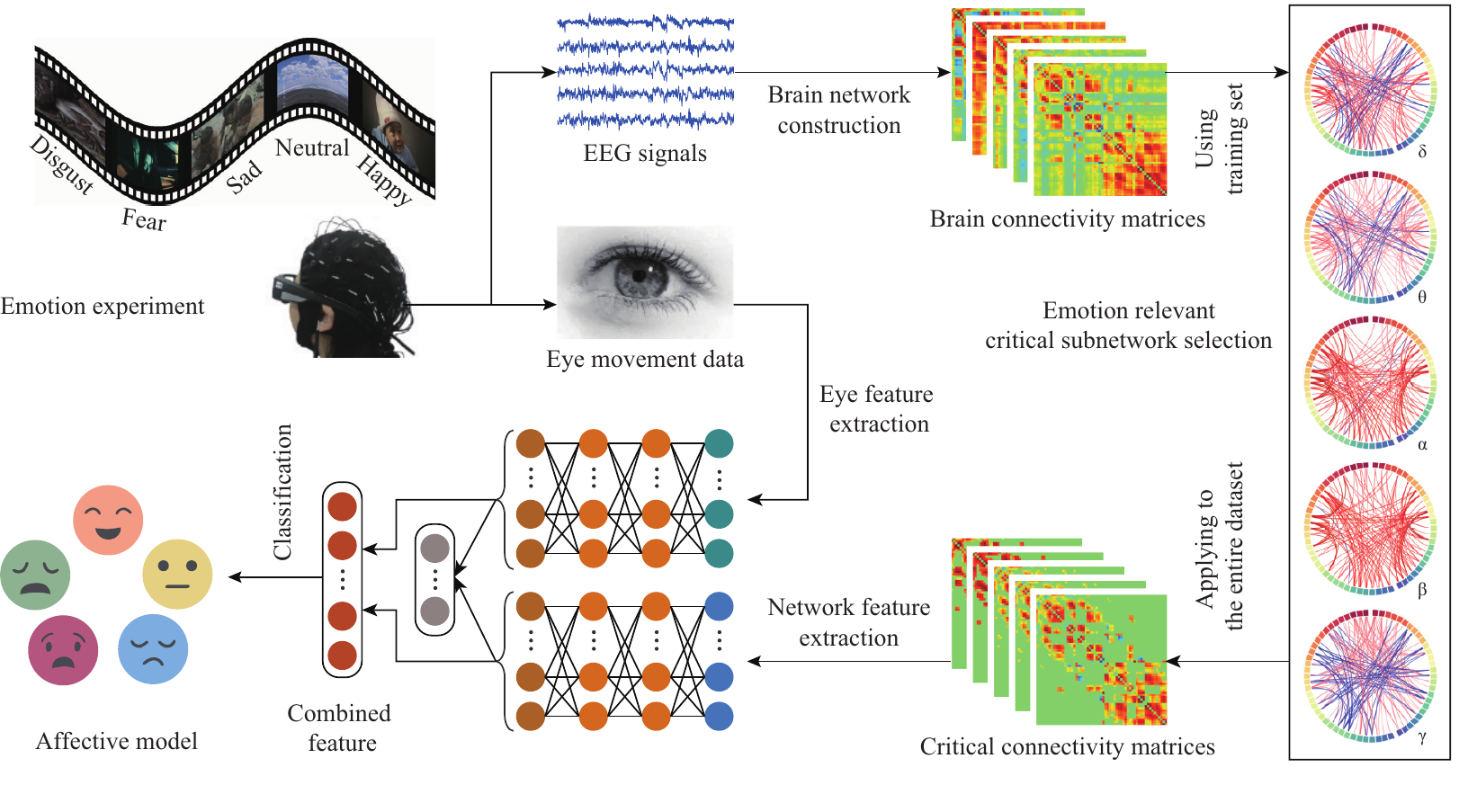}
	\caption{The framework of our proposed multimodal emotion recognition model using EEG-based functional connectivity and eye movement data. First, the emotion experiment is designed to simultaneously collect the EEG and eye tracking data while the subjects are watching emotional movie clips. Second, the EEG-based functional connectivity networks are constructed and selected to obtain the emotion-relevant critical subnetworks. Finally, the EEG functional connectivity network features and eye movement features are extracted and later fused to train the affective model.}
	\label{framework}
\end{figure*}

Among the physiological modalities, EEG has exhibited outstanding performance in emotion recognition and is promising in elucidating the basic neural mechanisms underlying emotion \cite{zheng2017identifying} \cite{garcia2019review}. Moreover, the fusion of EEG and eye tracking data has been shown efficient in multimodal emotion recognition, with increasing interests among research communities \cite{lu2015combining} \cite{lopez2016method}. In this paper, we adopt the EEG signals, along with eye movement data or peripheral physiological signals, to classify different emotions.

Most existing studies on EEG-based emotion recognition have relied on single-channel analysis \cite{zheng2017identifying, garcia2019review, lu2015combining, lopez2016method}, where EEG features are independently extracted within each EEG channel in different brain regions. In contrast, studies on cognitive science and neuroimaging have demonstrated that emotion is a complex behavioral and physiological reaction that involves circuits in multiple cerebral regions \cite{mauss2009measures}. In addition, studies in neuroscience and neuropsychiatry have revealed that patients with cognitive defect psychophysiological diseases such as autism, schizophrenia and major depressive disorder present decreased brain functional connectivity by both functional magnetic resonance imaging (fMRI) and EEG \cite{zhan2014deficient}. Furthermore, studies on neuroimaging based on fMRI have indicated that brain functional connectivity may offer the potential of representing the fingerprints in profiling individuals \cite{finn2015functional}, as well as the ability of individuals to sustain attention \cite{smith2016linking}. These results have provided evidence for the connection between cognition and brain functional connectivity. However, few studies have explored the emotion associated brain functional connectivity patterns. The study of emotion recognition from the perspective of the brain functional connectivity network remains to be further investigated and may eventually lead to the understanding of the underlying neurological mechanisms behind how emotions are processed in the brain.

In this paper, we aim to investigate the emotion-relevant brain functional connectivity patterns and evaluate the performance of the EEG connectivity feature for multimodal emotion recognition with respect to three public datasets: SEED \cite{lu2015combining}, SEED-V \cite{li2019classification}, and DEAP \cite{koelstra2011deap}. Fig. \ref{framework} depicts our proposed multimodal emotion recognition framework. The main contributions of our work lie in the following aspects:
\begin{enumerate}
	\item We propose a novel emotion-relevant critical subnetwork selection algorithm and investigate three EEG connectivity features: strength, clustering coefficient, and eigenvector centrality. 
	\item We demonstrate the outstanding performance of the EEG connectivity feature and its complementary representation properties with eye movement data in multimodal emotion recognition.
	\item We reveal the emotion associated brain functional connectivity patterns and the potential of applying the brain networks based on fewer EEG electrodes to aBCI systems in real scenarios.
\end{enumerate}

The remainder of this paper is organized as follows. Section II introduces the related literature regrading multimodal emotion recognition and brain functional connectivity analysis. Section III describes the emotion experimental design. Section IV presents the proposed multimodal emotion recognition framework based on the brain functional connectivity. Section V analyzes and discusses the experimental results. Finally, a brief conclusion will be presented in Section VI.

%
%
%
%
%
%

\section{Related Work}
\subsection{Emotion Recognition}
Various modalities have been exploited to detect affective states in the past few decades. With the advent of computer vision and speech recognition, research on emotion recognition using facial expression and speech has gained prevalence \cite{ko2018brief} \cite{schuller2018speech}. Hasani and Mahoor \cite{hasani2017facial} proposed an enhanced neural network architecture that consists of a 3D version of the Inception-ResNet network followed by a long short-term memory (LSTM) unit for emotion recognition from facial expressions in videos. They employed four databases in classifying different emotions, including anger, fear, disgust, sadness, neutrality, contempt, happiness, and surprise. Trigeorgis \textit{et al.} \cite{trigeorgis2016adieu} presented an end-to-end speech emotion recognition framework created by combining the convolutional neural network (CNN) and LSTM models. Schirmer and Adolphs \cite{schirmer2017emotion} studied different modalities with respect to emotion perception, including facial expression, voice, and touch. The authors suggested that the mechanisms of these modalities have their own specializations and that together, they could lead to holistic emotion judgment. 

Apart from the external behavioral modalities, the internal physiological signals have also attracted the attention of numerous researchers due to their objectivity and reliability. Zhang \textit{et al.} \cite{zhang2017respiration} conducted respiration-based emotion recognition using the sparse autoencoder and logistic regression model. Nardelli \textit{et al.} \cite{nardelli2015recognizing} performed valence-arousal emotion recognition based on the heart rate variability derived from the ECG signals. Atkinson and Campos \cite{atkinson2016improving} improved the EEG-based emotion recognition by combining the mutual information-based feature selection methods with kernel classifiers. Liu \textit{et al.} \cite{liu2017real} constructed a real-time movie-induced emotion recognition system to continuously detect the discrete emotional states in the valence-arousal dimension. Zheng \textit{et al.} \cite{zheng2017identifying} systematically evaluated the performances of different feature extraction, feature smoothing, feature selection and classification models for EEG-based emotion recognition. Their results indicated that stable neural patterns do exist within and across sessions. Among these modalities, EEG has been proven to be promising for emotion recognition and demonstrates competence for revealing the neurological mechanisms behind emotion processing.


In EEG-based emotion recognition, numerous EEG features have been exploited to enhance the performance of aBCI systems. The conventional EEG features could be categorized into temporal domain, frequency domain, and time-frequency domain \cite{wang2014tac}. In the temporal domain, the most commonly used EEG features mainly include the fractal dimension and higher order crossings \cite{jenke2014feature}. Due to the nonstationary essence of the EEG signals and the fact that raw EEG signals are usually contaminated with artifacts and noises, the frequency domain features such as power spectral density (PSD) \cite{zheng2015investigating}, higher order spectra \cite{jenke2014feature}, and differential entropy (DE) \cite{duan2013differential} and the time-frequency domain features such as wavelet features \cite{wang2014tac} and Hilbert-Huang spectra \cite{jenke2014feature} \cite{ackermann2016eeg} have demonstrated outstanding performance in the EEG-based emotion recognition systems. However, these conventional EEG feature extraction methods are based on single-channel analysis, which neglects the EEG-based functional connectivity networks in association with different emotions.

\subsection{Brain Functional Connectivity}
Brain connectivity has long been studied in the fields of neuroscience and neuroimaging to explore the essential nature of the cerebrum. According to the attributes of connections, brain connectivity could be classified into three modes: structural connectivity, functional connectivity, and effective connectivity \cite{rubinov2010complex}. These modes separately correspond to the biophysical connections between neurons or neural elements, the statistical relations between anatomically unconnected cerebral regions, and the directional causal effects from one neural element to another. 

Recently, increasing evidence has indicated that a link does exist between brain functional connectivity and multiple psychophysiological diseases with cognitive deficiency. Murias \textit{et al.} \cite{murias2007resting} found that robust patterns of EEG connectivity are apparent in autism spectrum disorders in the resting state. Yin \textit{et al.} \cite{yin2017functional} concluded that the EEG-based functional connectivity in schizophrenia patients tends to be slower and less efficient. Ho \textit{et al.} \cite{ho2015emotion} indicated that adolescent depression typically relates to the inflexibly elevated default mode network connections based on fMRI. Whitton \textit{et al.} \cite{whitton2018eeg} suggested that elevations in high frequency EEG-based functional connectivity may represent a neural pattern for the recurrent illness course of major depressive disorder. However, few studies have investigated the links between emotions and brain functional connectivity networks or conducted emotion recognition from the perspective of brain networks. Whether there truly exist specific connectivity patterns for different affective states remains to be lucubrated.

In the past years, only a few preliminary research efforts on EEG-based emotion recognition have attempted to employ the multichannel EEG analysis approaches. Dasdemir \textit{et al.} \cite{dasdemir2017analysis} directly used the connectivity metric of phase locking value as the EEG feature in distinguishing the positive and negative emotions. Lee and Hsieh \cite{lee2014classifying} tested three different connectivity metrics, correlation, coherence, and phase synchronization index, in classifying the positive, neutral, and negative emotions. Li \textit{et al.} \cite{li2019eeg} also studied these three emotions by combining the functional connectivity with local action features. Moon \textit{et al.} \cite{moon2018convolutional} utilized CNN to model the connectivity matrices constructed by three different connectivity metrics: correlation, phase locking value, and phase lag index. However, these studies either ignored the topology of the brain functional connectivity networks or failed to analyze the emotion-related functional connectivity signatures. In our previous study on EEG-based emotion recognition \cite{wu2019identifying}, we identified the brain functional connectivity patterns of the three emotions (sad, happy and neutral) and extracted the topological features from the brain networks to recognize these emotions. In this paper, we extend this preliminary work to the three-class (sad, happy, and neutral), five-class (disgust, fear, sad, happy, and neutral), and valence-arousal dimension multimodal emotion recognition tasks.

\subsection{Eye Movement Data}
Studies in neuroscience and biological psychology have indicated the relation between emotion and eye movement data, especially pupil diameter and dilation response. Widmann \textit{et al.} \cite{widmann2018emotion} indicated that emotional arousal by novel sounds is reflected in the pupil dilation response and the P3 event-related potentials. Oliva and Anikin \cite{oliva2018pupil} suggested that the pupil dilation response reveals the perception of emotion valence and confidence in the decision-making process. Moreover, Black \textit{et al.} \cite{black2017mechanisms} showed that the eye tracking and EEG data in autism spectrum disorders are atypical during the processes of attention to and cognition of facial emotions. 

In addition, eye movement data could be obtained through eye tracking glasses which are wearable, portable and noninvasive. Therefore, eye movement data, as a behavioral reaction to emotions, have been widely utilized to assist with EEG-based emotion recognition in aBCI systems. L\'{o}pez-Gil \textit{et al.} \cite{lopez2016method} improved EEG-based emotion recognition by combining eye tracking and synchronized biometrics to detect the valence-arousal basic emotions and a complex emotion of empathy. Zheng \textit{et al.} \cite{zheng2014multimodal} evaluated the complementary characteristics of EEG and eye movement data in classifying positive, neutral and negative emotions by fusing the DE and pupil diameter features. Lu \textit{et al.} \cite{lu2015combining} extended this preliminary work and systematically examined sixteen different eye movement features. Furthermore, their work has been extended to the five emotions by Li \textit{et al.} \cite{li2019classification} and Zhao \textit{et al.} \cite{zhao2019classification}, and the discrimination ability and stability over time of EEG and eye tracking data were also revealed. However, these research approaches were all based on single-channel analysis for the EEG signals: whether there exist complementary representation properties of EEG connectivity features and eye movement data remains to be further analyzed.

\subsection{Multimodal Frameworks}
As a complex psychological state, emotion is reflected in both physical behaviors and physiological activities \cite{adolphs2018neuroscience} \cite{mauss2009measures}. The collection of external behavioral data is more convenient than that of internal physiological signals, since the procedure could be accomplished without involving any invasive devices. Despite the inconvenience of data collection, the physiological signals are believed to be more objective and reliable because the participants cannot forge their internal activities. 

With different modalities exhibiting distinct properties, modern emotion recognition approaches have the tendency of combining multiple modalities to enhance the performance of aBCI systems. Perez-Gaspar \textit{et al.} \cite{perez2016multimodal} extended the evolutionary computation of artificial neural networks and hidden Markov models in classifying four emotions (angry, sad, happy, and neutral) by combining the speech with facial expressions. Tzirakis \textit{et al.} \cite{tzirakis2017end} also fused the auditory and visual modalities using an end-to-end valence-arousal emotion recognition model. They applied the CNN and ResNet models to extract features from speech and visual signals, respectively, which were then concatenated and fed into the LSTM model to accomplish the end-to-end training manner. Ranganathan \textit{et al.} \cite{ranganathan2016multimodal} conducted a 23-class discrete emotion recognition task based on four different deep belief networks by combining a variety of modalities, including face, gesture, voice and physiological signals. Huang \textit{et al.} \cite{huang2017fusion} studied the fusion of EEG and facial expression data using two decision-level fusion strategies, the sum and production rules, in detecting the four basic emotions (fear, sad, happy, and neutral).

In recent years, many researchers have suggested that the combination of EEG and eye tracking data is a promising approach for recognizing emotions in aBCI systems. L\'{o}pez-Gil \textit{et al.} \cite{lopez2016method} combined EEG with eye tracking and biometric signals in a synchronized manner to classify emotions using multiple machine learning methods. Liu \textit{et al.} \cite{liu2016emotion} applied the bimodal deep autoencoder (BDAE) neural network in detecting the three basic emotions (positive, neutral, and negative) from EEG and eye movement data. Tang \textit{et al.} \cite{tang2017multimodal} conducted the same task using bimodal deep denoising autoencoder and bimodal-LSTM models. Zheng \textit{et al.} \cite{zheng2018emotionmeter} presented EmotionMeter for detecting the four emotions (fear, sad, happy, and neutral). Qiu \textit{et al.} \cite{qiu2018multi} adopted the deep canonical correlation analysis (DCCA) model as a multimodal deep neural network for classifying the three-class, four-class, and valence-arousal emotions. Their results suggested that DCCA outperforms BDAE and bimodal-LSTM models in multimodal emotion recognition. In this paper, we apply the DCCA model to address the multimodal emotion recognition task.

\section{Emotion Experiment Design}
\subsection{Stimuli}
The emotion experiments were designed to simultaneously record the EEG and eye movement signals of the five prototypical emotions (disgust, fear, sad, happy, and neutral). Many existing works have indicated the efficiency and reliability of movie clips in eliciting the subjects' emotions due to the blending of audio and visual information \cite{lu2015combining} \cite{zheng2018emotionmeter}. Therefore, the movie clips were selected as the type of stimuli to better induce the subjects' affective states. 

During the preliminary experiment, a stimuli pool containing emotional movie clips corresponding with the five emotions was prepared and then assessed by 20 volunteers using rating scores ranging from 0 to 5. The higher scores represented the more successful elicitation of the subjects' emotions. Eventually, 9 movie clips for each of the five emotions were selected from the stimuli pool, all of which received a mean score of 3 or higher. The durations of these clips range from 2 to 4 minutes. 

\subsection{Subjects}
Sixteen subjects (6 males and 10 females) with normal hearing and self-reported normal or corrected-to-normal vision were recruited for our emotion experiments. All subjects were selected using the Eysenck Personality Questionnaire (EPQ), which could measure the personality of an individual in three independent dimensions: Extroversion/Introversion, Neuroticism/Stability, and Psychoticism/Socialization \cite{eysenck1985revised}. Those with extroverted characteristics and stable mood are more readily induced to experience the intended emotions throughout the experiment in comparison with those of other personalities. Hence, subjects that are more appropriate for the emotion experiments were selected according to the EPQ feedback.

\subsection{Protocol}
\begin{figure}[t!]
	\centering\includegraphics[scale=0.77]{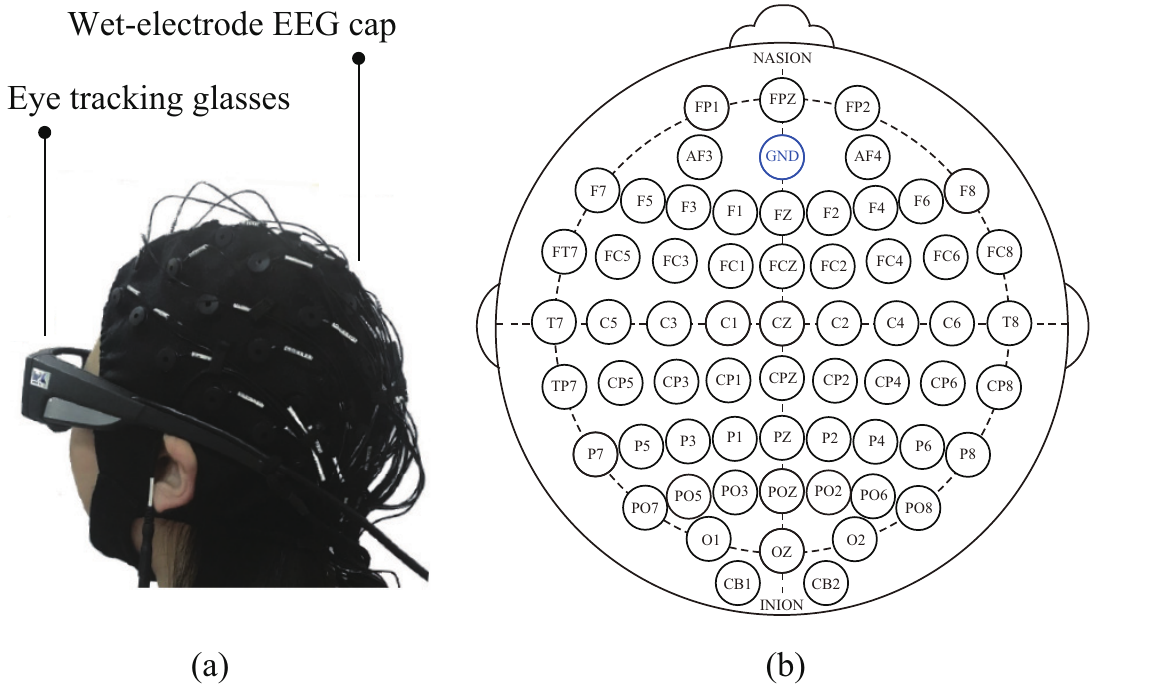}
	\caption{The wearable devices and the layout of the 62-channel EEG cap are presented in (a) and (b), respectively. Here, the GND channel denotes the reference electrode and is excluded from the 62 channels.}
	\label{device2}
\end{figure}

The emotion experiments were conducted under the laboratory environment. During the emotion experiment, the subject was required to view the emotional movie clips and relaxed as much as possible to induce their emotions. Meanwhile, their EEG and eye movement signals were simultaneously collected by the 62-channel wet-electrode cap and the SMI eye tracking glasses, respectively. The EEG data were recorded with the ESI NeuroScan System at a sampling rate of 1000 Hz. The layout of the 62-channel EEG cap is based on the higher-resolution international 10-20 system. Fig. \ref{device2} presents these wearable devices and the layout of the 62-channel EEG cap.

In this paper, there were 15 trials in total in each experiment, where each of the five emotions corresponds to 3 movie clips. Moreover, each subject was required to perform three sessions of the experiment on different days with an interval longer than three days. To better elicit the subjects' emotions, there was no repetition of movie clips within or across the three sessions. Thus, the aforementioned 9 movie clips for each emotion were randomly divided into three groups and later constructed the three sessions. As studies in cognitive science have indicated that emotion varies in a fluent and smooth manner, the order of play of these movie clips in one experiment was elaborately designed according to the following criteria: 1) avoiding sudden changes in emotion, such as clips of the emotion of disgust followed by clips of happiness; 2) utilizing movie clips of neutral emotion as a cushion between two opposite emotions. 

\begin{figure}[t!]
	\centering\includegraphics[scale=0.88]{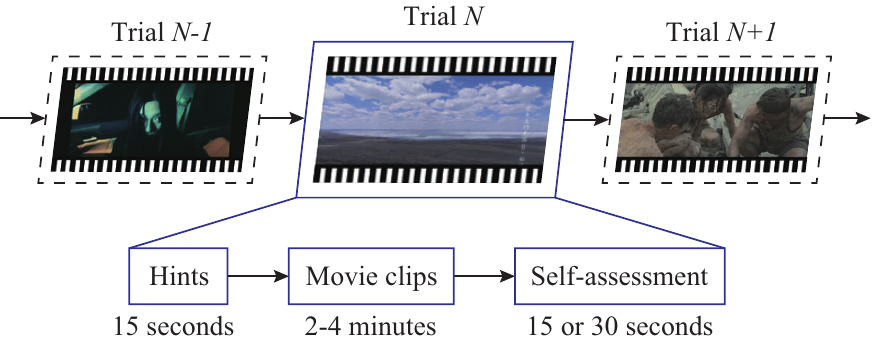}
	\caption{The protocol of the designed emotion experiment.}
	\label{protocol}
\end{figure}

Fig. \ref{protocol} illustrates the protocol of the designed emotion experiment. During each trial of the experiment, the movie clip was guided by 15 seconds of a brief introduction about the content and the emotion to be elicited and ended with 15 or 30 seconds of self-assessment and relaxation for the subjects to mitigate their emotions. Particularly, the resting time was 30 seconds after disgust or fear emotions, and 15 seconds after the other three emotions. In general, the duration of one experiment was approximately 55 minutes.

This multimodal emotion dataset is named SEED-V \cite{li2019classification}, which is a subset of the public emotion EEG dataset SEED\footnote{\url{http://bcmi.sjtu.edu.cn/~seed/}}. The SEED dataset \cite{lu2015combining} contains 62-channel EEG signals and eye movement data corresponding to three emotions (sad, happy, and neutral) from 9 subjects, with each subject performing the experiments three times. Thus, there are 27 experiments in total and each experiment contains 15 trials, with 5 movie clips for each of the three emotions.  

\subsection{Ethic Statement}
The emotion experiments have been approved by the Scientific \& Technical Ethics Committee of the Bio-X Institute at Shanghai Jiao Tong University. All of the subjects participating in our experiments were informed of the experimental procedures and signed the informed consent document before the emotion experiments.

\section{Methodology}
\subsection{Preprocessing}
The raw EEG signals collected during the emotion experiments are usually of high resolution and contaminated by surrounding artifacts, which hampers both the processing and the analysis of the emotion-relevant brain neural activities. To remove the irrelevant artifacts, the raw EEG data were preprocessed with Curry 7 to conduct the baseline correction, and a bandpass filter between 1 and 50 Hz was applied. Then, we downsampled the EEG signals to 200 Hz to expedite the processing procedures. For further exploration of the frequency-specific brain functional connectivity patterns, the EEG data were filtered with the five bandpass filters corresponding to the five frequency bands ($\delta$: 1-4 Hz, $\theta$: 4-8 Hz, $\alpha$: 8-14 Hz, $\beta$: 14-31 Hz, and $\gamma$: 31-50 Hz).

For the eye movement signals, the artifacts were eliminated using signals recorded from the EOG and FPZ channels. It has been proven that pupil diameter is subject to the ambient luminance in addition to the emotion stimuli materials \cite{bradley2008pupil}. Fortunately, according to our observation and analysis, the pupil diameter exhibits consistency in response to the same emotional stimuli material across different subjects. Thus, principal component analysis was adopted to eliminate the luminance reflex of the pupil and to preserve the emotion-relevant components.

\subsection{Brain Functional Connectivity Network}
The EEG-based brain functional connectivity networks consist of vertices and edges, which could be represented by the EEG electrodes and the associations between pairs of EEG signals from two different channels, respectively \cite{bullmore2009complex}. 

\subsubsection{Vertex Selection}
Although the wet-electrode EEG cap was adopted in our emotion experiments within the laboratory environment due to its reliability, the dry-electrode device with fewer EEG channels offers great convenience and portability in developing aBCI systems under actual scenario conditions. Thus, the question of whether the brain functional connectivity networks comprised of fewer EEG channels could exhibit considerable performance in emotion recognition remains unexplored. 

\begin{figure}[t!]
	\centering\includegraphics{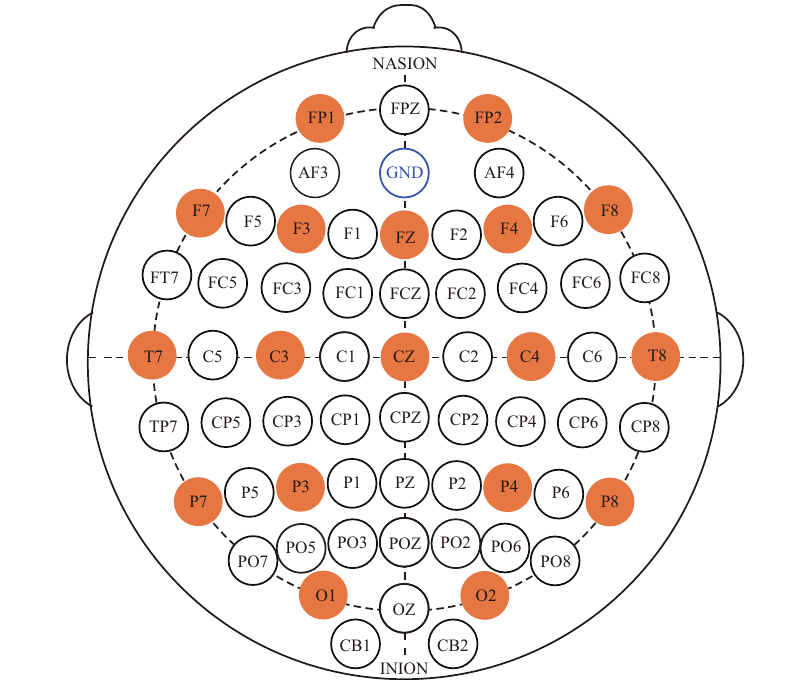}
	\caption{The locations of the 18-channel dry-electrode reflected in the layout of the 62-channel wet-electrode device. Here, the 18 electrodes are highlighted.}
	\label{10-20-2}
\end{figure}

Previous studies have demonstrated that the DSI-24 wearable sensing EEG headset is quite portable and appropriate for real scenarios \cite{tong2018sleep}. There are in total 18 channels in this device, and the layout of these electrodes is based on the normal international 10-20 system. Fortunately, these 18 electrodes can be perfectly mapped into the same locations in the layout of the higher-resolution international 10-20 system. Fig. \ref{10-20-2} displays the locations of the 18-channel dry-electrode reflected in the layout of the 62-channel wet-electrode device. 

In this paper, since the raw EEG signals were acquired with the 62-channel wet-electrode device, we constructed the brain connectivity network with 62 vertices in total.
Furthermore, we selected these 18 electrodes and compared the performances of EEG connectivity features extracted from the brain networks constructed with two categories of vertices: 18-channel and 62-channel.

\subsubsection{Edge Measurement}
To measure the associations between pairs of EEG signals recorded from different channels, we compared two connectivity metrics in this paper: Pearson's correlation coefficient and spectral coherence. The former can measure the linear relation between two EEG signals $x$ and $y$, which is defined as:
\begin{equation}
\rho_{x,y}=\frac{\text{cov}(x, y)}{\sigma_x\sigma_y}\label{corr},
\end{equation}
where $\text{cov}(x, y)$ denotes the covariance between $x$ and $y$, and $\sigma_x$ and $\sigma_y$ are the respective standard deviations. 

Distinguished from the correlation that measures the connectivity between two EEG channels in the temporal domain, coherence can measure the connectivity between two signals $x$ and $y$ at frequency $f$ in the frequency domain, which could be written as:
\begin{equation}
C_{x,y}(f)=\frac{|P_{xy}(f)|^2}{P_{xx}(f)P_{yy}(f)}\label{cohe},
\end{equation}
where $P_{xy}(f)$ is the cross power spectral density between $x$ and $y$, and $P_{xx}(f)$ and $P_{yy}(f)$ are the respective power spectral densities. 

\begin{figure}[t!]
	\centering\includegraphics[scale=0.8]{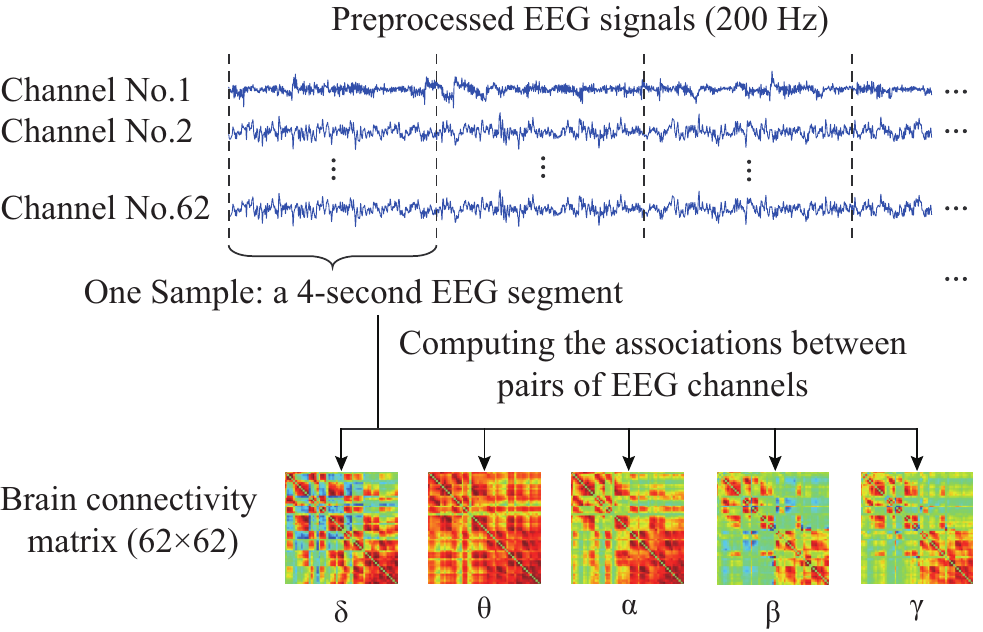}
	\caption{Illustration of the procedure for constructing EEG-based brain functional connectivity networks.}
	\label{construction}
\end{figure}

\subsubsection{Network Construction}
In our previous work, we have revealed the complementary characteristics of the state-of-the-art DE feature and eye movement data in emotion recognition, where features were extracted with a 4-second nonoverlapping time window \cite{lu2015combining} \cite{zhao2019classification} \cite{zheng2018emotionmeter} \cite{liu2019multimodal}. Considering further fusion of our proposed EEG connectivity feature with eye movement data and comparison with the DE feature, we constructed the brain functional connectivity network using the same time window. The detailed procedure of brain network construction is depicted in Fig. \ref{construction}. 

First, the preprocessed EEG signals were segmented with a 4-second nonoverlapping time window. As a result, each sample is represented by a 4-second EEG segment of 62 channels. Since the five bandpass filters were employed during the preprocessing, each sample actually contains five 4-second EEG segments corresponding to the five frequency bands. 

Second, the associations between pairs of EEG channels were computed using the connectivity metric (correlation or coherence) in each frequency band for each sample. Therefore, each brain network corresponds to a $62\times62$ symmetric connectivity matrix, where elements denote the association weights between pairs of EEG channels. Since the value of self-correlation is always equal to 1, elements on the main diagonal of connectivity matrices are usually set to zero and will not be used in later analysis \cite{rubinov2010complex}. 

Finally, five brain connectivity networks were acquired for each sample, corresponding to the five frequency bands. For the brain networks constructed with 18 channels, there is no need to repeat the above procedures. We could directly select the corresponding elements in the $62\times62$ connectivity matrix to construct the $18\times18$ connectivity matrix.

\subsection{Emotion-Relevant Critical Subnetwork Selection}
Although the raw EEG signals were preprocessed to remove the noises, there still remain certain minor artifacts that may not be eliminated during the preprocessing phase. Unfortunately, these artifacts may further lead to the weak associations in the brain networks, which eventually results in obscureness in profiling the brain network topology. By convention, this problem is resolved by directly discarding these weak associations according to an absolute or a proportional threshold after sorting the association weights \cite{rubinov2010complex} \cite{bullmore2009complex}. However, this method fails to take the targeted task into consideration, thus offering no guarantee that the preserved stronger associations are truly task relevant. Therefore, we have proposed an emotion-relevant critical subnetwork selection approach to address this issue.

The goal is to explore the universal emotion-relevant brain functional connectivity patterns among different subjects. Therefore, we utilized samples in training sets of all subjects to select the emotion-relevant critical subnetworks. Nevertheless, it should be mentioned that the affective models trained in this paper are subject-dependent. For further analysis of the brain connectivity patterns in different frequency bands, we selected a total of five critical subnetworks corresponding to the five frequency bands. Assuming that $L$ is the set of emotion labels, there are thus $|L|$ categories of emotions. The emotion-relevant critical subnetwork selection approach for one specific frequency band is summarized into the following three phases: 
\begin{enumerate}
	\item Averaging phase: all brain networks in training sets are averaged over all samples and all subjects for each emotion: thus, $|L|$ averaged brain networks corresponding to $|L|$ emotions could be obtained.
	\item Thresholding phase: for each averaged brain network, the same proportional threshold is applied to solely preserve the strongest associations. Thus, we attain critical edges for each emotion.
	\item Merging phase: along with the original vertices, the critical edges in all $|L|$ averaged brain networks are merged together to construct the emotion-relevant critical subnetwork.
\end{enumerate}
Here, the proportional threshold represents the proportion of the preserved connections relative to all connections in the brain network. Particularly, the threshold value was tuned as a hyperparameter of the affective models, ranging from [0.0, 1.0] with step size of 0.01. 

This procedure is also described in Algorithm \ref{alg:subnet}. Suppose that the connectivity matrices $X$ along with corresponding labels $Y$ in the training sets for one frequency band are defined as:
\begin{equation}
X=\{x^i\}_{i=1}^M\ (x^i\in\mathbb{R}^{N\times N}),\ Y=\{y^i\}_{i=1}^M\ (y^i\in L), 
\end{equation}
where $M$ and $N$ represent the number of samples and the number of vertices, respectively.

\begin{algorithm}[t!] 
	\caption{The proposed emotion-relevant critical subnetwork selection algorithm in a frequency band} 
	\label{alg:subnet} 
	\begin{algorithmic}[1] 
		\REQUIRE
		Connectivity matrices $X$, labels $Y$, label set $L$, threshold $t$, and vertices $V$
		\ENSURE
		The emotion-relevant critical subnetwork $G^*$
		\FOR{each $c\in L$}
		\STATE Average matrices with the same emotion label $c$: 
		\begin{equation*}
		\setlength{\abovedisplayskip}{1pt}
		\setlength{\belowdisplayskip}{1pt}
		x_c=\textit{mean}_{y^i==c}(x^i)
		\end{equation*}
		\STATE Sort the upper triangular elements of matrix $x_c$ based on the absolute value of the association weights: 
		\begin{equation*}
		\setlength{\abovedisplayskip}{1pt}
		\setlength{\belowdisplayskip}{1pt}
		x_c^*=\textit{sort}(\textit{abs}(\textit{triu}(x_c)))
		\end{equation*}
		\STATE Derive the critical edges corresponding to the strongest associations using threshold $t$:
		\begin{equation*}
		\setlength{\abovedisplayskip}{1pt}
		\setlength{\belowdisplayskip}{1pt}
		E_c =\textit{index}(x_c^*(1:t*N*(N-1)/2))
		\end{equation*}
		\ENDFOR
		\STATE Merge the critical edges of all emotions:
		\begin{equation*}
		\setlength{\abovedisplayskip}{1pt}
		\setlength{\belowdisplayskip}{1pt}
		E^* = \textit{union}_{c\in L}(E_c)
		\end{equation*}
		\STATE Construct the emotion-relevant critical subnetwork: 
		\begin{equation*}
		\setlength{\abovedisplayskip}{1pt}
		\setlength{\belowdisplayskip}{1pt}
		G^* = (V, E^*)
		\end{equation*}
		\RETURN $G^*$ 
	\end{algorithmic}
\end{algorithm}

\subsection{Feature Extraction}
\subsubsection{EEG Functional Connectivity Network Features}
In this paper, we extracted EEG features from the perspective of brain functional connectivity networks. The essence of the emotion-relevant critical subnetwork primarily consists in the network topology. According to the five critical subnetworks in the five corresponding frequency bands, we could derive the critical connectivity matrices for each sample in the entire dataset. Precisely, if one edge belongs to the critical subnetwork, the corresponding association weight in the matrix will remain unmodified; otherwise, it will be set to zero, thus simulating the process of discarding this edge from the brain network. The critical connectivity matrices were subsequently fed into the Brain Connectivity Toolbox \cite{rubinov2010complex} to extract the three topological features: strength, clustering coefficient, and eigenvector centrality.

Assume that the selected emotion-relevant brain functional connectivity network of one sample is regarded as an indirect graph $G=(V, E^*)$, where $V$ and $E^*$ represent the sets of vertices and critical edges, respectively. There are in total $N$ vertices in the brain network. Suppose that the corresponding symmetric connectivity matrix of $G$ is $A=(a_{ij})$, where $i,j=1,2,...,N$, $a_{ij}$ denotes the association between two vertices $v_i$ and $v_j$, and $a_{ij}=a_{ji}$. According to \cite{rubinov2010complex}, we could provide rigorous definitions for the three EEG functional connectivity network features as below.

The strength feature $F_S$ is a basic measurement of the network topology, which could be written as:
\begin{equation}
F_S=
\begin{bmatrix}
s_{i+}\big|_{i=1}^N, 
s_{i-}\big|_{i=1}^N,  
\sum\limits_{i=1}^N s_{i+}, \sum\limits_{i=1}^N s_{i-}
\end{bmatrix},
\end{equation}
where $s_{i+}$ and $s_{i-}$ represent the sum of the positive and negative associations connected to vertex $v_i$, respectively, and are computed as:
\begin{equation}
s_{i+}=
\sum\limits_{
	a_{ij}>0\atop
	j=1,2,..., N} a_{ij}, 
\end{equation}
\begin{equation}
s_{i-}=
\sum\limits_{
	a_{ij}<0\atop
	j=1,2,..., N} a_{ij}.
\end{equation}

The clustering coefficient feature $F_C$ is a measurement of the brain functional segregation that primarily quantifies the clusters within the brain network, which is defined as:
\begin{equation}
F_C=
\begin{bmatrix}
c_{i+}\big|_{i=1}^N, 
c_{i-}\big|_{i=1}^N,  
\sum\limits_{i=1}^N c_{i+}, \sum\limits_{i=1}^N c_{i-}
\end{bmatrix},
\end{equation}
where $c_{i+}$ and $c_{i-}$ represent the clustering coefficient vector for the positive and negative associations of vertex $v_i$, respectively. The clustering coefficient is equivalent to the fraction of triangles around a vertex and is calculated as: 
\begin{equation}
c_{i+}=
\dfrac{2t_{i+}}{k_i(k_i-1)},	
\end{equation}
\begin{equation}
c_{i-}=
\dfrac{2t_{i-}}{k_i(k_i-1)},
\end{equation}
where $k_i$ denotes the total numbers of neighbors for vertex $v_i$, and $t_{k+}$ and $t_{k-}$ are the positive and negative weighted geometric means of triangles around $v_i$, respectively. The triangles around a vertex are represented as:
\begin{equation}
t_{i+}=
\sum\limits_{j,h\in M_i\atop a_{ij},a_{ih},a_{jh}>0}
(a_{ij}a_{ih}a_{jh})^{1/3},
\end{equation}
\begin{equation}
t_{i-}=
\sum\limits_{j,h\in M_i\atop a_{ij},a_{ih},a_{jh}<0}
(a_{ij}a_{ih}a_{jh})^{1/3},
\end{equation}
where $M_i=\{v_j|e_{ij}\in E^*\}$ is the neighborhood of vertex $v_i$.

The eigenvector centrality feature $F_E$ evaluates the significance of an individual vertex in interacting with other vertices, facilitating integration, and thus serving a crucial role in network resilience. The eigenvector centrality score of vertex $v_i$ could be defined as:
\begin{equation}
F_E(i)=\dfrac{1}{\lambda}\sum\limits_{k\in M_i} F_E(k),
\end{equation}
where $M_i$ is the neighborhood of $v_i$. This equation could be easily transformed to the eigenvector equation using the vector notations: $\textbf{A}F_E=\lambda F_E$. In general, the dimensions of the strength, clustering coefficient, and eigenvector centrality features in each frequency band are $2N+2$, $2N+2$, and $N$, respectively. 

\subsubsection{Eye Movement Features}
First, eye movement parameters were calculated using the BeGaze\footnote{\url{https://gazeintelligence.com/smi-software-download}} analysis software of the SMI eye tracking glasses, including pupil diameter, fixation duration, blink duration, saccade, and event statistics. Subsequently, the statistics of these eye movement parameters were derived, thus obtaining the 33-dimensional eye movement feature. The detailed description of the extracted eye movement feature could be found in our previous work \cite{lu2015combining} \cite{zhao2019classification}.

\subsection{Classification}
As aforementioned, the variation of emotion is fluent and smooth, which should be reflected in the attributes of the extracted features. In addition, the extracted EEG features are typically of high dimensionality and may contain unrelated and redundant information, which increases the unnecessary computation and time costs. Hence, a feature smoothing method, linear dynamical system (LDS) \cite{duan2012eeg}, and a feature selection algorithm, minimal redundancy maximal relevance (mRMR) \cite{peng2005feature}, were applied to tackle this issue before feeding features into the classifier. 

\subsubsection{Deep Canonical Correlation Analysis Model}
\begin{figure}[t!]
	\centering\includegraphics[scale=0.9]{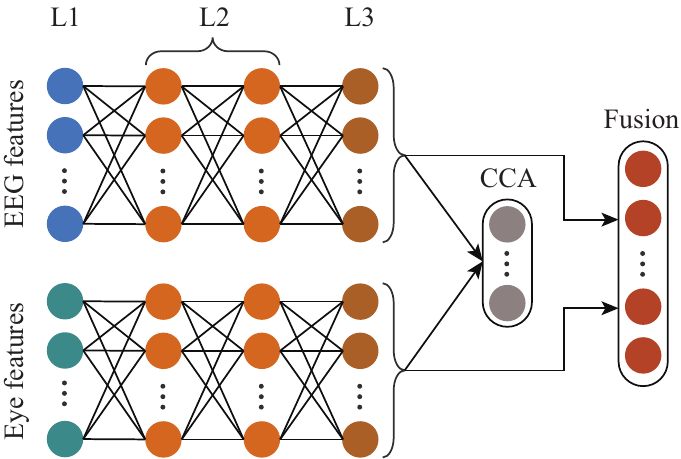}
	\caption{Architecture of the DCCA model.}
	\label{dcca}
\end{figure}

Fig. \ref{dcca} presents the architecture of the DCCA model, which comprises three parts: the stacked nonlinear layers (L2 and L3), CCA calculation, and feature fusion layer. The DCCA model could learn shared representations of high correlation from multimodal data \cite{andrew2013deep}. 

Assume that the transformed features for two modalities $X_1$ and $X_2$ are separately denoted by $H_1 = f_1(X_1; \theta_1)$ and $H_2 = f_2(X_2; \theta_2)$, where $f_1$ and $f_2$ are the respective nonlinear transformations, and $\theta_1$ and $\theta_2$ are the corresponding parameters. Thus, the optimization function is written as:
\begin{equation}
(\theta_1^*, \theta_2^*) 
 = \mathop{\arg\max}_{(\theta_1, \theta_2)} corr\big(f_1(X_1; \theta_1), f_2(X_2; \theta_2)\big).
\end{equation}
Suppose that the centered data matrices are $\bar{H_1}$ and $\bar{H_2}$, and $r_1$ and $r_2$ are the respective regularization parameters: hence, the correlation of the transformed features could be calculated as:
\begin{equation}
corr(H_1, H_2) = ||T||_{tr} = tr(T^\mathbb{'}T)^{1/2}, 
\end{equation}
where
\begin{equation}
\begin{split}
T & =\hat{\Sigma}_{11}^{-1/2}\hat{\Sigma}_{12}\hat{\Sigma}_{22}^{-1/2}, \\
\hat{\Sigma}_{11} & = \frac{1}{m-1}\bar{H_1}\bar{H_1}\mathbb{'}+r_1I, \\
\hat{\Sigma}_{22} & = \frac{1}{m-1}\bar{H_2}\bar{H_2}\mathbb{'}+r_2I, \\
\hat{\Sigma}_{12} & = \frac{1}{m-1}\bar{H_1}\bar{H_2}\mathbb{'}.
\end{split}
\end{equation}

\begin{table*}[t!]
	\caption{Mean accuracy (\%) and standard deviation (\%) of the EEG connectivity features in classifying the five emotions with respect to the SEED-V dataset.}
	\label{eegfea}
	\renewcommand{\arraystretch}{1.2}
	\begin{center}
		\begin{tabular}{c|c|c|c|c|c|c|c|c}
			\hline
			\hline
			Vertices & \multicolumn{4}{c|}{62-Channel} & \multicolumn{4}{c}{18-Channel} \\
			\hline
			Edges & \multicolumn{2}{c|}{Correlation} & \multicolumn{2}{c|}{Coherence} & \multicolumn{2}{c|}{Correlation} & \multicolumn{2}{c}{Coherence}  \\
			\hline
			\diagbox{Features}{Stat.} & Mean & Std. & Mean & Std. & Mean & Std. & Mean & Std.\\
			\hline
			Strength              		&	\textbf{74.05} & \textbf{7.09} 	&	71.89 & 7.20 	&	\textbf{72.63} & \textbf{8.26} 	&	71.46 & 6.08\\ \hline
			Clustering Coefficient      &	56.35 & 11.14 	&	71.18 & 7.82 	&	51.94 & 8.63 	&	64.14 & 6.10\\ \hline
			Eigenvector Centrality 		&	68.78 & 9.39 	&	60.80 & 7.18 	&	67.89 & 11.13 	&	65.57 & 7.38\\		
			\hline
			\hline
		\end{tabular}
	\end{center}
\end{table*}

In particular, the gradient of $corr(H_1, H_2)$ could be computed using singular value decomposition. The parameter updating is accomplished by using the negative value of correlation as the loss function. Thus, minimizing loss is equivalent to maximizing correlation. The feature fusion layer is defined as the weighted average of the two transformed features \cite{qiu2018multi}. Finally, the fused multimodal feature is fed into the support vector machine (SVM) to train the affective model. 

In this paper, the cross validation and grid search methods were adopted to tune the hyperparameters. Supposing that the numbers of nodes in L1, L2, and L3 layers of DCCA are $n_1$, $n_2$, and $n_3$, respectively, these three hyperparameters are searched in the space where $n_1\ge n_2 \ge n_3$ and $n_1, n_2, n_3 \in \{2^5, 2^6, 2^7, 2^8\}$. The learning rate is tuned from $10^{-8}$ to $10^{-4}$. 

\subsubsection{Experiment Setups}
In this paper, we evaluate the proposed approaches on three public datasets: SEED \cite{lu2015combining}, SEED-V \cite{li2019classification}, and DEAP \cite{koelstra2011deap}. For the SEED dataset, the three-class (sad, happy, and neutral) emotion classification task is conducted. The training and test sets are the first 9 trials and the last 6 trials, respectively, which is the same as in \cite{lu2015combining} \cite{liu2016emotion} \cite{tang2017multimodal} \cite{liu2019multimodal}. For the SEED-V dataset, the five-class (disgust, fear, sad, happy, and neutral) emotion classification task is performed with a three-fold cross validation strategy, which follows the same setups as in \cite{zhao2019classification} \cite{liu2019multimodal}. 

The DEAP dataset contains 32-channel EEG signals and 8-channel peripheral physiological signals from 32 subjects in the valence-arousal dimension. Each subject watched 40 one-minute music videos. The EEG signals were preprocessed with a bandpass filter between 4 and 45 Hz. For the DEAP dataset, we build the brain networks using solely 32 channels in the four frequency bands (without the $\delta$ band) with a 2-second nonoverlapping time window. The peripheral physiological feature is 48-dimensional. In addition, two binary (arousal-level, valence-level) classification tasks were conducted with ten-fold cross validation strategy. The setups for the DEAP dataset are in accordance with \cite{liu2016emotion} \cite{tang2017multimodal} \cite{liu2019multimodal}.

\section{Experimental Results and Discussion}
\subsection{Discrimination Ability}
To demonstrate the discrimination ability of the EEG functional connectivity network features in emotion recognition, we conduct EEG-based emotion recognition for the three datasets.

\subsubsection{Experimental Results on the SEED-V Dataset}

For the SEED-V dataset, we constructed the EEG-based brain functional connectivity networks using two different categories of vertices and two different edge measurements, then extracted three EEG functional connectivity network features from the brain networks. 

Table \ref{eegfea} presents the five-class emotion recognition performance of these features. We could observe that the strength feature exhibits outstanding performance regardless of the number of vertices and connectivity metric. This may be because the strength feature could intuitively reflect the emotion associated connectivity of the entire brain regions. In general, the strength and eigenvector centrality features exhibit higher accuracy with correlation as the connectivity metric, whereas the clustering coefficient feature exhibits better performance with coherence.

The features extracted from 18-channel-based brain networks exhibit considerable performance compared with those of 62-channel networks, which indicates that the EEG functional connectivity network features extracted from the brain networks constructed with fewer channels are promising for actual scenarios of emotion recognition applications in aBCI systems. 

In our previous work \cite{wu2019identifying}, we have demonstrated that the EEG functional connectivity network features considerably outperform the PSD feature and that they are superior to those directly using the connectivity metrics as features. In this paper, the best classification accuracy of $74.05\pm7.09$\% achieved by the strength feature defeats the value of $69.50\pm10.28$\% attained by the single-channel-based state-of-the-art DE feature in the work of \cite{zhao2019classification} for the same dataset.

\begin{figure}[thpb]
	\centering\includegraphics[scale=0.93]{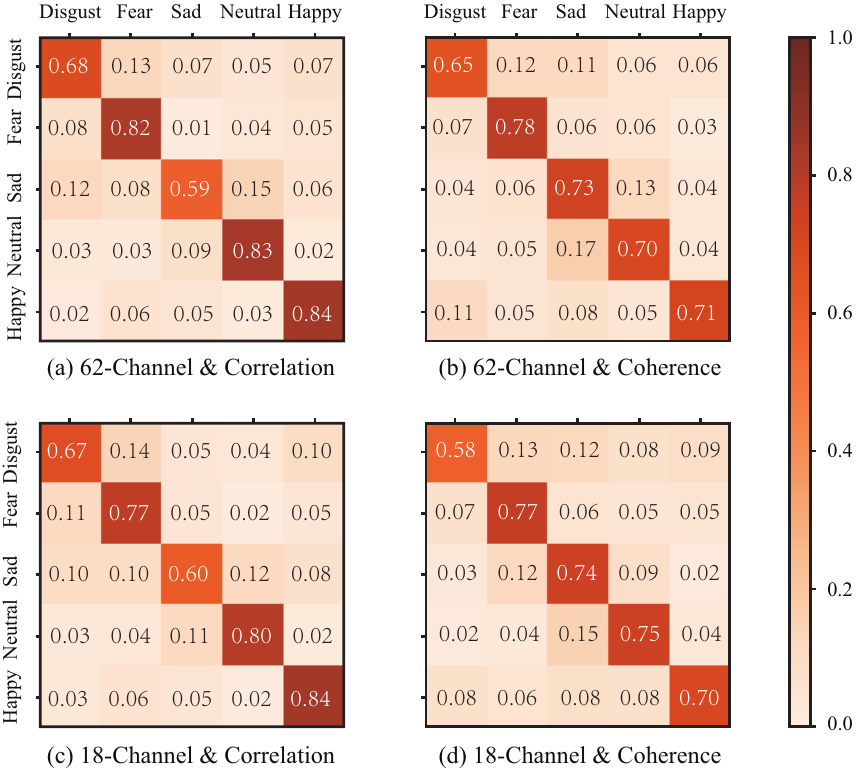}
	\caption{The confusion matrices of the strength feature extracted from the brain networks constructed with two categories of vertices and two connectivity metrics.}
	\label{strength4}
\end{figure}

To further analyze the capability of the best feature, the strength, in recognizing each of the five emotions, the confusion matrices are displayed in Fig. \ref{strength4}. It could be observed that the strength feature is superior in detecting the emotion of happiness, followed by the emotions of neutrality and fear with correlation as connectivity metric. In contrast, from the perspective of coherence as connectivity metric, the strength feature exhibits the best performance in recognizing the fear emotion and exhibits similar performances with respect to the emotions of sadness, neutrality, and happiness. 

Generally, the strength feature with correlation as connectivity metric could achieve better performance in classifying all emotions, except sadness, in comparison with that of coherence. Overall, the EEG feature exhibits fair performance in recognizing the emotion of disgust and could be easily confused by the sad and neutral emotions, the results for which are in accordance with previous findings \cite{zhao2019classification}. In addition, the strength feature with the 18-channel approach exhibits considerable performance compared with that of the 62-channel approach. Particularly, the strength feature with the 18-channel approach could achieve the same classification accuracy of 84\% with that of the 62-channel approach with respect to recognizing the emotion of happiness.

\subsubsection{Experimental Results on SEED and DEAP Datasets}
On SEED and DEAP datasets, we utilize the best EEG functional connectivity network feature, the strength, to further verify its discrimination ability in classifying emotions.

On the SEED dataset, the three-class emotion recognition accuracy achieved by the strength feature is $80.17\pm7.12\%$, which is higher than the value of $78.51\pm14.32\%$ \cite{lu2015combining} attained by the DE feature. On the DEAP dataset, the performances of the strength feature for two binary classification tasks (arousal-level and valence-level) are $73.42\pm4.67\%$ and $76.10\pm4.49\%$, respectively. These results considerably outperform those of $62.0\%$ and $57.6\%$ \cite{koelstra2011deap} achieved by the PSD feature, as well as those of $68.28\%$ and $66.73\%$ \cite{chao2019sensor} attained using the capsule network.

These results demonstrate the discrimination ability of the EEG functional connectivity network features in classifying three-class emotions (sad, happy, and neutral), five-class emotions (disgust, fear, sad, happy, and neutral), and valence-arousal dimension. Additionally, the strength feature outperforms the most commonly used PSD feature and the state-of-the-art DE feature.

\subsection{Complementary Representation Properties}
In this section, the DCCA model is adopted to combine the EEG signals with other modalities for multimodal emotion recognition with respect to the three datasets. Here, the best EEG functional connectivity network feature, strength with correlation as connectivity metric, is utilized for the evaluation.

\subsubsection{Experimental Results on the SEED-V Dataset}
The combination of EEG and eye movement data for the SEED-V dataset was implemented using two different fusion strategies: feature-level fusion (FLF) and DCCA. The FLF is a direct concatenation of the two modalities' features. The experimental results are displayed in Table \ref{multimodal}. The best classification performance values (\%) based on the EEG connectivity feature, eye movement data, FLF, and DCCA approaches are 74.05$\pm$7.09, 65.21$\pm$7.60, 78.03$\pm$6.07, and 84.51$\pm$5.11, respectively. These results indicate that the combination of the EEG connectivity feature and eye movement data could enhance the performance of five-class emotion recognition. Moreover, the DCCA model may find the shared space to be more related to emotion. In addition, the fusion based on the 18-channel EEG connectivity feature and eye movement data also achieves considerable classification performance. 

\begin{table}[thpb]
	\caption{Performance (\%) of two single modalities and two multimodal fusion strategies in classifying the five emotions with respect to the SEED-V dataset.}
	\label{multimodal}
	\renewcommand{\arraystretch}{1.2}
	\centering
	\begin{threeparttable}
		\begin{tabular}{c|c|c|c|c}
			\hline
			\hline
			Vertices & \multicolumn{2}{c|}{62-Channel} & \multicolumn{2}{c}{18-Channel} \\ 
			\hline
			Stat. & Mean & Std. & Mean & Std.  \\ \hline
			EEG & 74.05 & 7.09 & 72.63 & 8.26 \\ \hline
			EYE & 65.21 & 7.60 & 65.21 & 7.60 \\ \hline
			FLF & 78.03 & 6.07 & 78.02 & 7.30 \\ \hline
			DCCA & \textbf{84.51} & \textbf{5.11} & \textbf{84.45} & \textbf{6.10} \\
						
			\hline\hline
		\end{tabular}		
	\end{threeparttable}
\end{table}

Table \ref{multimodal2} presents the performance of our proposed EEG feature compared with the single-channel-based state-of-the-art DE feature \cite{zhao2019classification} \cite{liu2019multimodal} for the multimodal emotion recognition task with respect to the SEED-V dataset. These results reveal that our proposed EEG connectivity feature outperforms the DE feature in combination with eye movement data to classify the five emotions, whether using the 62-channel or 18-channel-based functional connectivity networks. 

\begin{table}[thpb]
	\caption{Classification performance (\%) of different works in multimodal emotion recognition on the SEED-V dataset.}
	\label{multimodal2}
	\renewcommand{\arraystretch}{1.2}
	\centering
	\begin{threeparttable}
		\begin{tabular}{c|c|c|c}
			\hline
			\hline
			Works & Method & Mean & Std. \\ \hline
			
			\multirow{2}{*}{Zhao \textit{et al.} \cite{zhao2019classification} }
			& FLF & 73.65 & 8.90 \\ \cline{2-4}
			& BDAE & 79.70 & 4.76 \\ \hline
			
			\multirow{3}{*}{Liu \textit{et al.} \cite{liu2019multimodal} }		
			& Max & 73.17 & 9.27 \\ \cline{2-4}
			& Fuzzy & 73.24 & 8.72 \\ \cline{2-4}
			& DCCA & 83.08 & 7.11 \\ \hline
			
			\multirow{2}{*}{Our method (62-channel)}
			& FLF & 78.03 & 6.07 \\ \cline{2-4}
			& DCCA & \textbf{84.51} & \textbf{5.11}\\ \hline
			
			\multirow{2}{*}{Our method (18-channel)}
			& FLF & 78.02 & 7.30 \\ \cline{2-4}
			& DCCA & \textbf{84.45} & \textbf{6.10} \\
			\hline\hline
		\end{tabular}		
	\end{threeparttable}
\end{table}

\begin{figure}[thpb]
	\centering\includegraphics[scale=0.93]{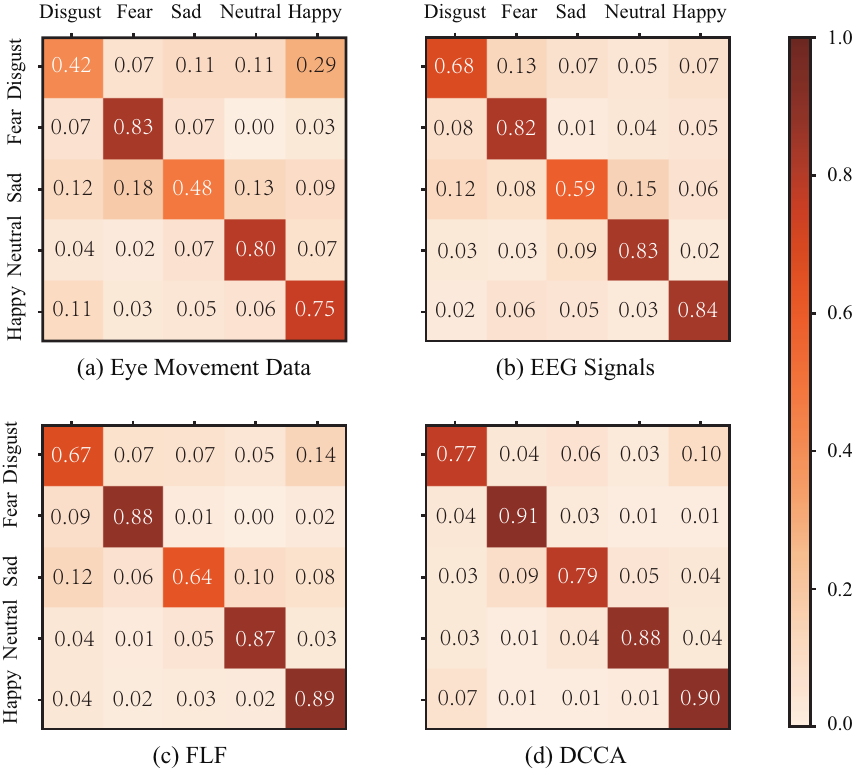}
	\caption{Confusion matrices of five-class emotion recognition using: (a) eye movement data, (b) EEG signals, (c) FLF, and (d) DCCA models.}
	\label{fusion4}
\end{figure}

To investigate the capabilities of EEG connectivity feature and eye movement data in detecting each specific emotion, the confusion matrices are displayed in Fig. \ref{fusion4}. Here, the EEG feature is the strength feature with 62 channels and the correlation connectivity metric. 

\begin{figure}[thpb]
	\centering\includegraphics{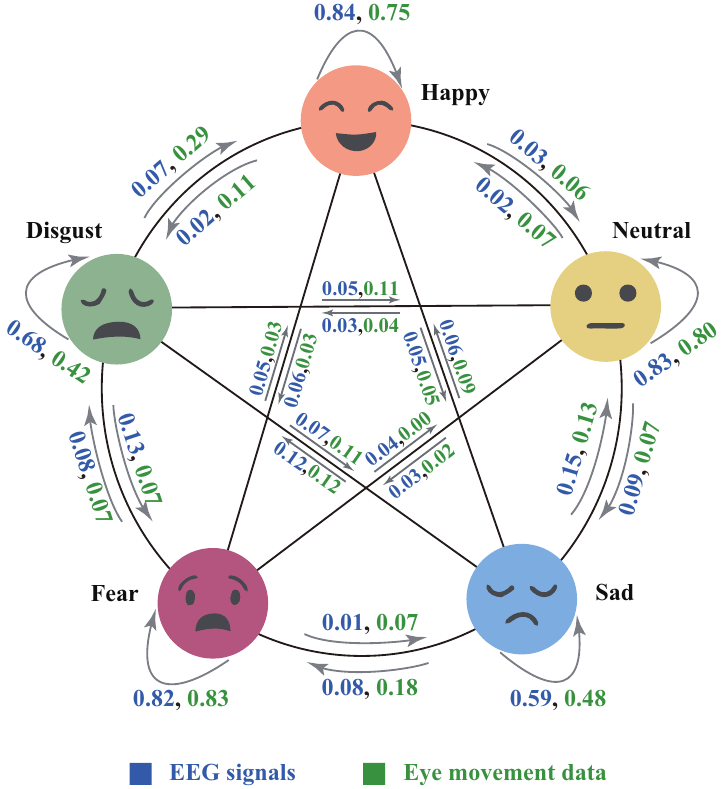}
	\caption{Confusion graph of the EEG functional connectivity network feature and eye movement data in classifying the five emotions: disgust, fear, sadness, happiness, and neutrality.}
	\label{emotion5}
\end{figure}

It could be observed that both EEG and eye movement data exhibit potential in classifying the emotions of fear, happiness, and neutrality. In particular, the EEG connectivity feature dominates the recognition of the happiness emotion, while eye movement data excel at detecting the fear emotion. The confusion graph of these two modalities is also presented in Fig. \ref{emotion5}. 

In comparison with the single modality affective model, the last two confusion matrices in Fig. \ref{fusion4} indicate that the multimodal fusion strategies could indeed improve the classification performance for all of the five emotions. These results demonstrate the complementary representation properties of the EEG connectivity feature and eye movement data in classifying the five emotions.

\subsubsection{Experimental Results on SEED and DEAP Datasets}
The classification performances of our work and several existing works with respect to the SEED dataset are displayed in Table \ref{seed3}. The existing works are based on the DE feature and eye movements. We could observe that the best performance of $95.08\pm6.42\%$ is achieved by our work, which combines the strength feature with eye movement data to detect three emotions (happiness, neutrality, and sadness). These results further verify that the combination of EEG and eye movements could enhance the classification performance.

\begin{table}[thpb]
	\caption{Classification performance (\%) of different works in multimodal emotion recognition with respect to the SEED dataset.}
	\label{seed3}
	\renewcommand{\arraystretch}{1.2}
	\centering
	\begin{threeparttable}
		\begin{tabular}{c|c|c|c}
			\hline
			\hline
			Works & Method & Mean & Std. \\ \hline
			
			\multirow{2}{*}{Lu \textit{et al.} \cite{lu2015combining}} 
			& FLF & 83.70 & - \\ \cline{2-4}
			& Fuzzy & 87.59 & - \\ \hline
			
			Song \textit{et al.} \cite{Song2018EEG} 
			& DGCNN & 90.40 & 8.49 \\ \hline
			
			Liu \textit{et al.} \cite{liu2016emotion}
			& BDAE & 91.01 & 8.91 \\ \hline
			
			Tang \textit{et al.} \cite{tang2017multimodal} 
			& Bimodal-LSTM & 93.97 & 7.03 \\ \hline

			Liu \textit{et al.} \cite{liu2019multimodal}
			& DCCA & 94.58 & \textbf{6.16} \\ \hline
			
			Our method & DCCA & \textbf{95.08} & 6.42  \\
			
			\hline\hline
		\end{tabular}		
	\end{threeparttable}
\end{table}

Table \ref{deap} presents the classification performances of our work and several existing works with respect to the DEAP dataset. The existing works are based on the combination of peripheral physiological features with the PSD \cite{xing2019fin} \cite{yin2017cmpb} or DE \cite{liu2016emotion} \cite{tang2017multimodal} \cite{liu2019multimodal} features. The highest classification accuracy of the two binary classification tasks, $85.34\pm2.90\%$ for the arousal-level and $86.61\pm3.76\%$ for the valence-level, are both obtained by our work. These results reveal that the strength feature is also superior to the PSD and DE features in fusion with peripheral physiological signals.

\begin{table}[thpb]
	\caption{Classification performance (\%) of different works in multimodal emotion recognition on the DEAP dataset.}
	\label{deap}
	\renewcommand{\arraystretch}{1.2}
	\centering
	\begin{threeparttable}
		\begin{tabular}{c|c|c|c|c|c}
			\hline
			\hline
			\multirow{2}{*}{Works}  & \multirow{2}{*}{Method} & \multicolumn{2}{c|}{Arousal} & \multicolumn{2}{c}{Valence}  \\ \cline{3-6}
			
			& & Mean & Std. & Mean & Std. \\ \hline
			
			Xing \textit{et al.} \cite{xing2019fin} 
			& SAE-LSTM & 74.38 & - & 81.10 & - \\ \hline	
			
			Liu \textit{et al.} \cite{liu2016emotion}
			& BDAE & 80.50 & 3.39 & 85.20 & 4.47 \\ \hline
								
			Tang \textit{et al.} \cite{tang2017multimodal}
			& Bimodal-LSTM & 83.23 & 2.61 & 83.82 & 5.01 \\ \hline
			
			Yin \textit{et al.} \cite{yin2017cmpb} 
			& MESAE & 84.18 & - & 83.04 & - \\ \hline
			
			Liu \textit{et al.} \cite{liu2019multimodal}
			& DCCA & 84.33 & \textbf{2.25} & 85.62 & \textbf{3.48}  \\ \hline
			
			Our method 
			& DCCA & \textbf{85.34} & 2.90 & \textbf{86.61} & 3.76  \\
			\hline\hline
		\end{tabular}		
	\end{threeparttable}
\end{table}

\subsection{Critical Frequency Bands}
In this section, we evaluate the critical frequency band of the EEG functional connectivity network feature on the SEED-V dataset. 

\begin{figure}[thpb]
	\centering\includegraphics[scale=0.9]{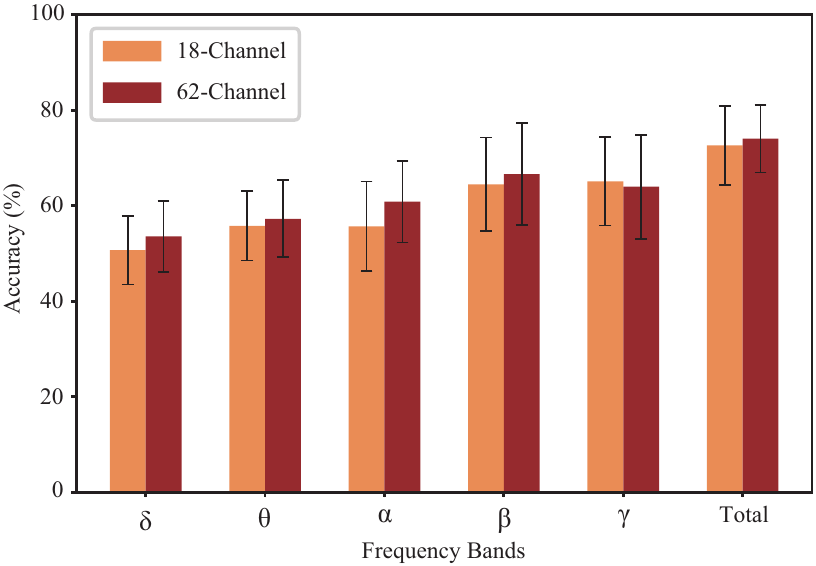}
	\caption{Classification performance (\%) of different frequency bands using the strength feature with two different categories of vertices.}
	\label{bandacc2}
\end{figure}

Fig. \ref{bandacc2} presents the classification performance of different frequency bands using the strength feature with correlation as the connectivity metric. The result demonstrates that the $\beta$ and $\gamma$ frequency bands are superior in classifying the five emotions in comparison with other bands, which is in accordance with the results attained by the DE feature \cite{zheng2015investigating} \cite{wu2019identifying}. Additionally, the frequency bands with the 18-channel approach achieve comparable performance with that of the 62-channel approach, which implies the possibility of applying 18 electrodes to detect emotions in real scenario applications.

\subsection{Brain Functional Connectivity Patterns}

\begin{figure*}[thpb]
	\centering\includegraphics[scale=0.9]{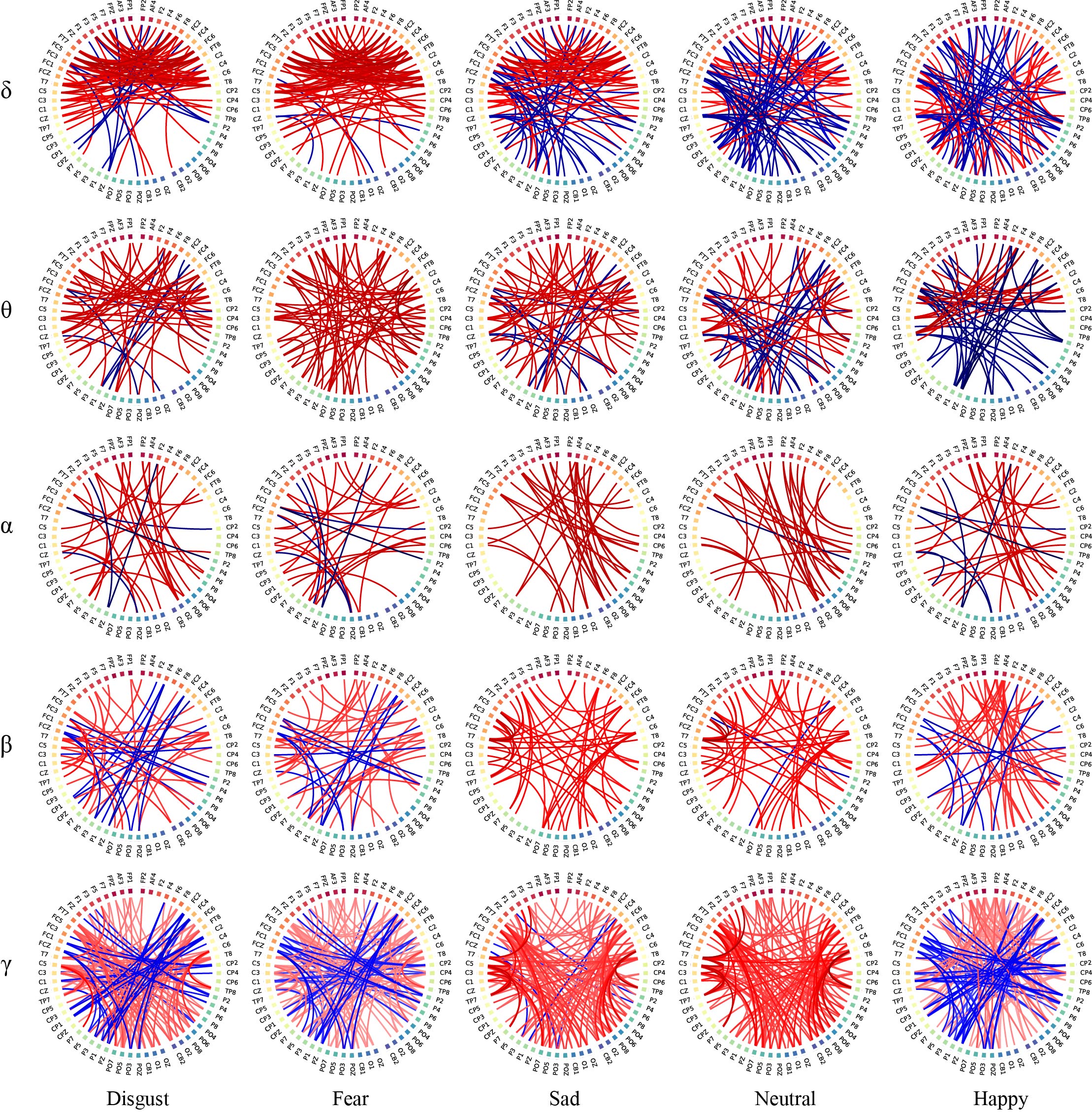}
	\caption{The brain functional connectivity patterns for the five emotions in the five frequency bands with correlation as connectivity metric. In each subfigure, the positive and negative associations are drawn in red and blue colors, respectively, with darker line colors reflecting larger absolute values of association weights. In addition, the 62 nodes in each circle denote the 62 EEG electrodes located according to the following criterion: 1) the electrodes in the left and right cerebral regions lie in the left and right parts of the circle, respectively; 2) the electrodes in the frontal, temporal, parietal, and occipital areas are displayed from the top to bottom of each circle.}
	\label{pattern}
\end{figure*}

In this section, we investigate the brain functional connectivity patterns based on the SEED-V dataset. The emotion-relevant critical subnetworks are selected through three phases: averaging, thresholding, and merging. The numbers of subnetworks attained after each phase are 25, 25, and 5, respectively. Specifically, 25 corresponds to the five emotions in the five frequency bands, while 5 refers to the five frequency bands, since the critical connections of the five emotions in each frequency band are merged together during the third phase. 

In this paper, the 25 subnetworks attained after the thresholding phase are adopted to analyze the frequency-specific brain functional connectivity patterns in association with the five emotions. To analyze both the positive and negative connections, we visualize the brain functional connectivity networks based on the connectivity metric of correlation.

Considering that the subnetworks are selected using samples from training sets of all participants and that a three-fold cross validation strategy is utilized, the emotion-relevant critical subnetworks are calculated three times. The results demonstrate that stable connectivity patterns are exhibited across these three calculations. 

Fig. \ref{pattern} presents the 25 critical subnetworks associated with the five emotions in the five frequency bands averaged over three folds. For better analysis of the distinct connectivity patterns for each emotion in each frequency band, we display all of the critical connections except for the intersections among the five emotions.

It could be observed that the positive correlation connectivity is much higher in the frontal lobes in the $\delta$ band for the negative affective states, including the emotions of disgust, fear, and sadness. In particular, for the disgust emotion, stronger positive connectivity is exhibited in the $\gamma$ bands within both the left and right brain regions, and stronger negative connectivity is observed between the left and right brain regions. However, the fear emotion in the $\gamma$ band is dominated by the stronger negative connectivity, and there are much weaker positive connections compared with those of the disgust emotion. In addition, much more positive connectivity in the $\theta$ band is exhibited for the fear emotion. 

The fact that the functional connectivity patterns are quite similar for the sad and neutral emotions could account for the confusion between the sad and neutral emotions, which is consistent with previous findings \cite{li2019classification} \cite{zheng2015investigating}. Nevertheless, the connectivity in the $\delta$ band tends to be positive within the frontal areas and negative in the left brain regions for the sadness emotion, while negative in larger brain areas for the neutral emotion. 

In terms of the happiness emotion, the entire cerebral areas are much more active in the $\delta$ band with both positive and negative correlation connectivity. Moreover, in the $\theta$ band, negative connectivity is revealed between the frontal and parietal lobes, with positive connectivity between the frontal and temporal lobes. In the $\gamma$ band, the functional connectivity patterns for the emotions of happiness, fear, and disgust are more similar, which may be originated from the fact that amygdala voxels contribute to these three emotions \cite{saarimaki2015discrete}. Overall, these results are in accordance with findings in the literature based on fMRI that the brain regions contributing to the emotion classification are predominated in the frontal and parietal lobes \cite{saarimaki2015discrete}.

%

\section{Conclusion}
In this paper, we have proposed a novel emotion-relevant critical subnetwork selection algorithm and evaluated three EEG connectivity features (strength, clustering coefficient, and eigenvector centrality) on three public datasets: SEED, SEED-V, and DEAP. The experimental results have revealed that the emotion associated brain functional connectivity patterns do exist. The strength feature is the best EEG connectivity feature and outperforms the state-of-the-art DE feature based on single-channel analysis. Furthermore, we have performed the multimodal emotion recognition using the DCCA model based on the EEG connectivity feature. The classification accuracies are $95.08\pm6.42\%$ on the SEED dataset, $84.51\pm5.11\%$ on the SEED-V dataset, and $85.34\pm2.90\%$ and $86.61\pm3.76\%$ on the DEAP dataset. These results have demonstrated the complementary representation properties between the EEG connectivity feature and eye movement data. Additionally, the results have indicated that the brain functional connectivity networks based on the 18-channel approach are promising for multimodal emotion recognition applications in aBCI systems under actual scenario situations.


%





\ifCLASSOPTIONcaptionsoff
  \newpage
\fi



%
\bibliographystyle{IEEEtran}
\bibliography{IEEEabrv,bibfile}

\end{document}